# Superconductivity in the vicinity of a ferroelectric quantum phase transition

S. E. Rowley[1,2,*], C. Enderlein[2,3,*], J. Ferreira de Oliveira[2], D. A. Tompsett[1], E. Baggio Saitovitch[2], S. S. Saxena[1,4,*] and G. G. Lonzarich[1,*]

1. Cavendish Laboratory, University of Cambridge, J. J. Thomson Avenue, Cambridge, CB3 0HE, United Kingdom

2. Centro Brasileiro de Pesquisas Físicas, Rua Dr Xavier Sigaud 150, Rio de Janeiro, 22290-180, Brazil

3. UFRJ, Estrada de Xerém 27, Xerém, Duque de Caxias, Rio de Janeiro, 25245-390, Brazil

4. National University of Science and Technology MISiS, Leninsky Prospekt 4, Moscow 119049, Russia

**Superconductivity has been observed in electron-doped $SrTiO_3$ at charge-carrier densities below $10^{18}$ per $cm^3$, where the density of states at the Fermi level of the itinerant electrons is several orders of magnitude lower than that of conventional metals. In terms of the Bardeen-Cooper-Schrieffer (BCS) description for a single electronic band, this implies the existence of an extraordinarily strong interaction driving the formation of Cooper pairs, potentially comparable in order of magnitude to that in some high-temperature superconductors. Under suitable conditions the interaction might remain effective at densities approaching metallic densities, leading to the possibility of Cooper pair formation at elevated temperatures. To shed further light on the mechanism for superconductivity in $SrTiO_3$ and related materials, we have investigated the pressure dependence of the electrical resistivity and superconducting transition temperature, $T_c$, at a carrier density near to the maximum of the superconducting dome in the temperature-carrier density phase diagram. Our experiments show that $T_c$ in metallic $SrTiO_3$ collapses rapidly with increasing pressure and hence with increasing frequency of the soft transverse optical polar phonon mode connected with the ferroelectric quantum critical point. We show that the phase diagram of $T_c$ vs. carrier density and vs. frequency of the soft polar phonon mode can be understood in terms of the coupling of charge carriers via two hybrid longitudinal polar modes, based on a physical model of long-wavelength dipolar fluctuations of the charge carrier-ion system as a starting point. In particular, the model predicts that for carrier densities above a threshold of the order of $10^{18}$ $cm^{-3}$, $T_c$ can be strongly enhanced on approaching the ferroelectric quantum critical point, qualitatively as observed in our measurements in $SrTiO_3$ and as found in many electrically conducting**

* E-mail addresses of key authors for correspondence/materials: ser41@cam.ac.uk, carsten@xerem.ufrj.br, sss21@cam.ac.uk, ggl1@cam.ac.uk

**magnetic analogues. However, below this threshold density we predict the reverse behaviour, namely that $T_c$ can be strongly suppressed on approaching the ferroelectric quantum critical point, in agreement with predictions of theoretical analyses presented recently based on controlled treatments of the Eliashberg theory. Our model is also relevant to superconductivity found in gated ferroelectric quantum critical systems such as $KTaO_3$ and is expected to guide searches for new superconductors in a diversity of materials.**

Strontium titanate is an incipient ferroelectric insulator, which can be tuned essentially continuously into the ferroelectric state via a 'quantum' tuning parameter such as chemical substitution[1,2], isotopic substitution[3,4] or applied strain[5,6]. The temperature-quantum tuning parameter phase diagram of $SrTiO_3$ and related materials has recently been discussed in terms of a phenomenological field-theoretic model including the interaction of the local ferroelectric order parameter field with itself, and with the strain field of the lattice[3,7-19]. A self-consistent perturbative treatment of the model in an isotropic approximation leads to a good quantitative description of the phase diagram and the temperature dependence of the inverse dielectric function, $1/\varepsilon_0$, without the use of free adjustable parameters[3]. The phase diagram is characterized by (i) a ferroelectric transition line that terminates at a quantum critical point, QCP, (ii) a low temperature crossover line separating a power law ($1/\varepsilon_0 \sim T^2$) and an exponential temperature dependence of $1/\varepsilon_0$ also terminating at the QCP, and (iii) a high temperature crossover line separating classical (Curie-Weiss) and quantum behaviour of the temperature dependence of $1/\varepsilon_0$. An additional crossover line that terminates at the QCP marks the position of the minimum in the temperature dependence of $1/\varepsilon_0$, which arises from the coupling of the ferroelectric order parameter to the lattice strain field[3,13,20-23]. Interestingly, in contrast to conventional expectations this coupling appears to be insufficient to lead to an experimentally detectable first order transition near the QCP in $SrTiO_3$. A more rigorous treatment of this coupling including lattice anisotropy however[14,17], predicts exotic behaviour beyond that indicated above[24] at least for sufficiently low frequencies and low temperatures below that probed experimentally thus far. These effects have also be observed in a number of other systems for example in oxides[25], organics[20], hydrogen-bonded crystals[24], electronic ferroelectrics[26] and multiferroics[21] and have excited interest in recent reviews[27,28].

The substitution of Ti by Nb leads to the introduction of an extra electron per substituted unit cell of $SrTiO_3$, which is bound to Nb only very weakly due to the high dielectric constant of the host lattice. For a doped electron density, $n$, above the order of $10^{15}$ cm$^{-3}$ the doped electrons are promoted to the $t_{2g}$ valence bands that are split by the spin-orbit interaction as well as by a weak perturbation observed below approximately 100 K of the starting simple cubic structure of $SrTiO_3$ [29]. The carriers (conduction electrons) occupying the lowest of these $t_{2g}$ bands tend to dominate the properties of principal interest in the discussion below, which will focus on relatively low carrier densities $n$ in the range $10^{17}$ to $10^{21}$ cm$^{-3}$. Umklapp processes are absent under these conditions as are intervalley scattering processes which would require the presence of multiple Fermi surface pockets well separated in the Brillouin zone, a fact not realised in electron-doped SrTiO$_3$ as confirmed by several quantum oscillation experiments and band structure calculations (see for example refs. (29) and (30) and also the Extended Data section). Cooper pair formation and superconductivity in such low carrier-concentration metals are seen by many investigators as surprising and remarkable even after half a century since its observation in SrTiO$_3$ as the first of the oxide superconductors[31]. Despite numerous investigations, the detailed nature of the relevant effective interaction mediating superconductivity in carrier-doped SrTiO$_3$ [31-54], and related materials such as KTaO$_3$ [55], and interface materials such as LaAlO$_3$/SrTiO$_3$ [47,56] and FeSe/SrTiO$_3$ [57], continues to be debated.

Here we present the pressure dependence of the superconducting transition temperature, $T_c$, as determined from resistivity measurements at a carrier density, $n$, near to the peak of the ambient-pressure superconducting dome in the temperature-carrier density phase diagram (i.e. close to optimal doping) of SrTiO$_3$ (Figs. 1a and 1b). As shown in these figures we find that $T_c$ drops sharply with modest pressures and collapses towards absolute zero above 5 kbar. This is consistent with a peak in $T_c$ coinciding with the minimum as a function of pressure of the ambient-pressure inverse dielectric function, $1/\varepsilon_0$ (Fig. 1c) as determined by dielectric susceptibility measurements, or alternatively the minimum of the soft transverse-optical polar phonon frequency[5,7-9,58,59], $\Omega(q)$, connected with the ferroelectric quantum critical point, which is only weakly dependent on $n$ particularly up to optimal doping[60,61]. Thus the superconducting dome appears to be centred in the vicinity of a ferroelectric QCP in a way reminiscent of the behaviour observed on the border of magnetic phase transitions in metallic systems[62]. This conclusion is further supported by recent measurements of increases of $T_c$ observed upon calcium[63] and oxygen-18 isotope[64] substitutions bringing electron-doped SrTiO$_3$ even closer to its quantum critical point, i.e. by further reducing $\Omega(q)$ at low temperature at $q = 0$. It is also supported by the observation of a lower value of $T_c$ in electron-doped KTaO$_3$ [55] which is a quantum paraelectric further away from the quantum critical point (having 1.5 times the value of $\Omega(q=0)$ at low temperature compared to SrTiO$_3$ in the undoped starting material[3]). Note that the 'ferroelectric' transition in the conducting state[65], where the uniform dielectric function is strictly singular for all finite $n$, is defined by the condition $\Omega(q) \rightarrow 0$ at small wavenumber $q$. As already intimated, the symbol $\varepsilon_0$ will denote throughout the starting uniform static dielectric function for the undoped state. The variation with pressure of the *A* coefficient of the resistivity (defined by $\rho = \rho_0 + AT^x$) is seen in Fig. 1d to vary by approximately 30% over the pressure range where $T_c$ drops by an order of magnitude or more. The higher sensitivity of $T_c$ than *A* to pressure is qualitatively consistent with the expected exponential dependence of $T_c$ vs. a power-law dependence of *A* on the electron-phonon coupling scale[66,67].

The above findings and our pressure measurements shown in Fig. 1a provide further constraints on models for the mechanisms for carrier pairing in these systems. In light of the proximity of the superconducting dome to the ferroelectric quantum critical point, it is natural to consider first the role of the polar optical modes consisting of a multiplet of transverse and longitudinal components, the lowest transverse mode frequency $\Omega(q)$ as defined above, vanishing at the ferroelectric transition at $q = 0$. Analyses suggest that for our purposes (see e.g., Ruhman & Lee[50]) it is sufficient to consider the highest longitudinal optical (LO) polar mode, $\Omega_{LO}$ at approximately 100 meV in the starting undoped state and weakly *q* dependent, and the lowest transverse optical (TO) polar mode, $\Omega(q)$, which can be one or more orders of magnitude lower than $\Omega_{LO}$, varying inversely with the square root of the uniform static dielectric susceptibility, $\varepsilon_0$, in the insulating state.

In terms of the dielectric screening model[44,68-72], the above effective pairing interaction in the low *n* limit is attractive in the frequency range $\Omega(q) < \omega < \Omega_{LO}$, which extends to $\omega = 0$ at low *q* when $\varepsilon_0$ diverges ($\Omega(q) \to 0$) at the ferroelectric critical point (model given below). In this limit the induced interaction arises from the coupling of the conduction electrons with essentially bare charges (i.e., the polar ions) leading to an effective coupling strength, analogous to a deformation potential, that can be much larger than that in the case of charge neutral atoms and in conventional phonon mediated superconductivity. At higher *n* the doped conduction electrons can neutralize the polar ions leading to a collapse of the attractive interaction and hence of $T_c$, in agreement with observation (inset of Fig. 1a). This collapse may not, however, be inevitable under all circumstances, and a deeper understanding of the above effective interaction may provide the insights needed to find ways to preserve a strong ionic coupling strength up to carrier densities well above that predicted by the present model.

It has been suggested that the mediation of polar optical modes, while of key importance in reducing the Coulomb repulsion as $\varepsilon_0$ increases, might be supplemented by an additional pairing mechanism to account for superconductivity or at least for a quantitative understanding of the magnitude of $T_c$. A number of additional candidates have been proposed involving: (i) plasmons in the conduction electron system[50,53], which must be included for minimal consistency of any proposed description, (ii) multi-valley transition processes[31,41], (iii) non-polar acoustic phonons[43], (iv) non-polar soft optical phonons[40], (v) currents associated with transverse polar optical modes, (vi) non-cancelling contributions of polar optical modes in between $\Omega(q)$ and $\Omega_{LO}$ [50,52], (vii) phonon modes localized around the dopant impurity sites[52] and (viii) effects associated with Sr disordering at low temperatures[73]. Models involving the formation of polarons and bipolarons[42], as well as pre-formed pairs[37], have also been considered. A number of studies have largely ruled out cases (ii)-(v) for SrTiO$_3$ but the importance of the remaining mechanisms continue to be debated.

Having carefully investigated all of these proposed theories in light of realistic model parameters for SrTiO$_3$ and our new experiments we reconsider here a minimal description that includes pairing due to the dipolar fluctuations of the coupled ion system (polar optical modes) and the charge-carrier system in a dielectric screening model of the effective long-wavelength interaction between carriers of the form

$$V(q,\omega) = \frac{4\pi e^2}{q^2 \varepsilon(q,\omega)} \quad (1)$$

where $\varepsilon(q,\omega)$ is the wavevector and frequency dependent dielectric function and $q$ and $\omega$ measure, respectively, the momentum and energy transfers in two-electron scattering processes. In these calculations we adopt units in which $\hbar = k_B = 1$. We assume that for each mode of wavevector $q$, $\varepsilon(q,\omega)$ can be described by simple harmonic oscillator contributions from fluctuations of the dipolar density of the ions and the carriers[74] as given approximately for low $q$ by two resonance terms

$$\frac{\varepsilon(q,\omega)}{\varepsilon_\infty} = 1 + \frac{\omega_p^2}{\omega(q)^2 - \omega^2} + \frac{\Omega_p^2}{\Omega(q)^2 - \omega^2} \quad (2a)$$

where $\omega_p$ and $\Omega_p$ are the bare plasma frequencies for the carriers and ions respectively in a medium with background dielectric constant $\varepsilon_\infty$,

$$\omega(q) = v_F q/\sqrt{3} \quad (2b)$$

is the bare characteristic frequency at low $q$ for density fluctuations in the carrier system with Fermi velocity $v_F$,

$$\Omega(q) = \sqrt{\Omega(0)^2 + v_s^2 q^2} \quad (2c)$$

is the bare frequency spectrum at low $q$ for ionic fluctuations characterized by a frequency gap, $\Omega(0)$ and a sound velocity, $v_s$. These are the two transverse mode frequencies where $\varepsilon(q,\omega)$ diverges and the interaction $V(q,\omega)$ vanishes. We note that the resonance terms associated with the charge carriers reduce to that due to Lindhard[72] both in the low and high $\omega/v_F q$ limits and provide an interpolation formula[50,52] in between for $q$ below the Fermi wavenumber, $k_F$.

The inverse of $\varepsilon(q,\omega)$ can also be expressed in terms of the sum of two resonance modes so that the effective interaction $V(q,\omega)$ defined in Eq. 1 can be rewritten using Eq. 2 in the form

$$V(q,\omega) = \frac{4\pi e^2}{q^2 \varepsilon_\infty}\left(1 - \frac{\gamma_- \omega_-(q)^2}{\omega_-(q)^2 - \omega^2} - \frac{\gamma_+ \omega_+(q)^2}{\omega_+(q)^2 - \omega^2}\right) \quad (3)$$

where $\omega_\pm(q)$ are the coupled carrier-ion longitudinal frequencies (hybrid longitudinal frequencies), defined by the condition $\varepsilon(q,\omega) \to 0$, and $\gamma_\pm(q)$ are the corresponding coupling parameters, all determined straightforwardly by the starting material parameters defining $\varepsilon(q,\omega)$ in Eq. (2). In contrast to the starting transverse modes, the hybrid longitudinal mode frequencies, $\omega_\pm(q)$, correspond to cooperative motion of the charge carriers and the ions. We postulate that these hybrid longitudinal modes are the dominant source of attraction between conduction electrons.

At a low carrier density, $n$, where $\omega_p \ll \Omega_p$ the lower hybrid longitudinal mode of frequency $\omega_-(q) < \omega_p$ can be thought of as a carrier plasma mode as screened by the ions, so that $\omega_-(q)$ can be much lower than $\omega_p$ (and even than $v_F k_F$, where $k_F$ is the Fermi wavenumber), while the upper hybrid longitudinal mode of frequency $\omega_+(q)$ can be described as a weakly renormalized longitudinal optical polar phonon mode. On the other hand at high carrier density, $n$, where $\omega_p \gg \Omega_p$ the lower hybrid longitudinal mode of frequency $\omega_-(q) < \omega_p$ can be thought of as a polar optical phonon mode as screened by the charge carriers, so that $\omega_-(q)$ tends to $\Omega(q)$, while the upper hybrid longitudinal mode of frequency $\omega_+(q)$ can be described as a weakly screened carrier plasma mode. The $n$ dependences of $\omega_\pm(q)$ are illustrated in Fig. 2a for starting parameters relevant to SrTiO$_3$ (Methods).

Equally important are the coupling functions $\gamma_-(q)$ and $\gamma_+(q)$ (Fig. 2b) that affect the resonance weights in Figs. 2c and 2d, which show the calculated real part of the normalized retarded interaction potential based on Eqs. 1-3. The coupling parameter $\gamma_-(q)$, corresponding to the lower hybrid longitudinal mode, tends to be suppressed by screening with increasing $\varepsilon_0$. At low $n$ ($\omega_p \ll \Omega_p$), where the upper hybrid longitudinal frequency $\omega_+(q)$ tends to be far above $v_F k_F$ (Fig. 2a) and hence ineffective in pairing, the suppression of $\omega_-(q)$ along with $\gamma_-(q)$ tends to reduce $T_c$ with increasing $\varepsilon_0$. At high $n$ ($\omega_p \gg \Omega_p$), where the upper hybrid longitudinal frequency $\omega_+(q)$ is not far above $v_F k_F$ (Fig. 2a) and hence potentially important in pairing, the suppression of $\gamma_-(q)$ and $\omega_-(q)$ tends to open up a wide area of attraction up to $\omega_+(q)$ (Fig. 2c) that can potentially enhance rather than suppress superconductivity with increasing $\varepsilon_0$, i. e., on approaching a ferroelectric QCP.

In an attempt to gain insight on the conditions favourable for superconductivity for the pairing interaction defined by Eqs. 1-3 we examine the predictions of the spin-singlet BCS[75] gap equation defining $T_c$ in the weak coupling approximation[76,77]

$$\Delta(k) = \sum_{k'} U(k,k') \frac{\Delta(k')}{2\xi(k')} tanh\left(\frac{\xi(k')}{2T_c}\right) \qquad (4)$$

where $\Delta(k)$ is the energy gap function, $\xi(k)$ is the one-electron energy spectrum measured from the Fermi level, and $U(k,k')$ is an appropriate interaction kernel to be defined below and in Methods. We assume that the band can be treated in an effective mass approximation with an effective mass, $m$, which in general depends on $n$.

It is sometimes assumed that the kernel can be approximated at least in the weak coupling limit by the interaction $V(q,\omega)$ with $q = k' - k$ and $\omega = \xi(k') - \xi(k)$ equal to the momentum and energy transfers, respectively, in a process in which a pair of electrons in one-particle states with wavevectors $k$ and $-k$ scatter into states with wavevectors $k'$ and $-k'$. However, this 'on-shell' approximation does not fully take into account the way in which the dynamics of the interacting charge carriers affect the total relevant interaction. For example, in the on-shell approximation the effective pairing interaction is independent of the velocities of the interacting carriers when the two velocities are equal, which is inconsistent with the intuitive expectation that the polarization of the medium produced by moving carriers should be dependent on their velocities. Also excluded are the effects of changes in vacuum fluctuations due to changes in the occupation of the one-particle states that accompany each pair transition $(-k, k) \rightarrow (-k', k')$. Such effects are normally described by the Eliashberg theory[78], which may be used to infer an improved approximation of the kernel $U(k,k')$ in Eq. (4). In the analysis of Kirzhnits, Maksimov, and Khomskii (KMK)[70], originally applied to the case of SrTiO$_3$ by Takada[44], the kernel $U(k,k')$ differs significantly from $V(q,\omega)$ but is nevertheless essentially fully specified by it. More precisely $U(k,k')$ is given by a particular average along the imaginary frequency axis of the continuation of $V(q,\omega)$ in the complex frequency plane (Methods). This formulation in terms of imaginary frequencies amounts to an elegant book-keeping procedure for dealing, for example, with the blocking or unblocking of vacuum fluctuations accompanying pair transitions as discussed above. Importantly $U(k,k')$ depends on the absolute values of the velocities of the interacting carriers and not on the velocity difference assumed in the on-shell approximation.

The KMK kernel, however, does not fully describe all of the physical effects included in the more computationally demanding and subtle Eliashberg theory[79,80]. Numerical analyses show that the KMK approximation more faithfully reproduces the predictions of the Eliashberg theory than the on-shell approximation[81].

However, it has been argued that a still more serious challenge to both the KMK and Eliashberg descriptions can arise form vertex corrections and potentially from spin fluctuation corrections in cases involving interactions mediated by bosons with energies well above the Fermi energy[79,82-84]. In light of these still poorly resolved effects we limit ourselves to consider only to what extent the model interaction, Eqs. 1-3, can account for the observed trend of $T_c$ vs. density and pressure. For this purpose we restrict ourselves to the KMK approximation that is more accurate than the on-shell approximation and more straightforward to apply and interpret than the full Elishberg formalism.

The gap equation, Eq. (4), can be represented in operator form as

$$\Lambda \, \Delta = K \, \Delta \qquad (5)$$

where $\Delta$ is a vector and $K$ is an operator whose matrix form is defined by Eq. (4) with $\Lambda \to 1$. This is an eigenvalue equation and the eigenvalues can be found in the standard way by evaluating the determinant of the matrix $K - \Lambda I$, where $I$ is the unit matrix. The transition temperature $T_c$ is found by the condition that the highest eigenvalue, $\Lambda_h$, is equal to unity.

Instead of determining $T_c$ directly we could consider the behaviour of $\Lambda_h$ in the low temperature limit as a function, in particular, of the carrier density and applied pressure. The region in density and pressure where superconductivity is expected to arise would be indicated by the condition $\Lambda_h > 1$. More generally, the maximum of $\Lambda_h$ vs. density and pressure may be expected to indicate the regime where the contribution of the pairing interaction defined by Eqs. 1-3 to the total pairing interaction is strongest.

The details of the evaluation of the highest eigenvalue based on Eqs. 1-5 together with parameters representative for SrTiO$_3$ are given under Methods. Instead of presenting $\Lambda_h$ vs. carrier density $n$ and pressure $p$ for a fixed low value of $T$, here we present results of the predicted $n$ and $p$ dependences of $T_c$ inferred from the condition $\Lambda_h(T = T_c, n, p)$ is equal to unity. This procedure depends on the known material properties plus one adjustable parameter, namely the cut-off frequency in the sum defining the gap equation. Importantly, the effect of this cut-off is comparatively weak for the trends vs. $n$ and $p$ of normalized values of $T_c$, and especially of normalized values of the highest eigenvalue $\Lambda_h$. The trends of $\Lambda_h$ in particular can hence be viewed as predictions of the model, Eqs. 1-5, that are essentially free of adjustable parameters.

We note that typical techniques used to calculate superconductivity involve determining parameters "$\lambda$" (connected to the attraction between electrons) and "$\mu$" (connected to the Coulomb repulsion of like charges). These effects are seamlessly incorporated into the model in the form of the interaction defined by Eqs. (1-3). The effects relating to the renormalization of "$\mu$" to "$\mu^*$" in typical calculations are taken into account here by the form of the interaction and the calculation used to determine the gap function in Eqs. (4-5). This renormalization process is physically related to the fact that one of the particles in a two-particle pairing process is able to avoid the repulsive core of the other particle by changing its position as a function of space and time (sometimes referred to as retardation). In particular in the case of superconductivity in SrTiO$_3$, the oscillations in time of the gap function and interaction function are synchronised such that electron repulsion may be avoided. In terms of the frequency dependent functions this means that the gap function as a function of frequency changes sign when the interaction function becomes positive (repulsive). This can therefore be thought of as the frequency-space analogue of the phenomenon occurring in $k$-space (wavevector space) in magnetically mediated d-wave superconductors in which the gap function changes sign in $k$-space to help avoid particle repulsion.

The predicted variations of the normalized values of $T_c$ vs. $n$ and $p$ are summarized in Fig. 3. From the inset we see that the calculated $T_c$ initially rises with carrier density, $n$, reaches a maximum in the range $10^{19}$ to $10^{20}$ cm$^{-3}$, and collapses at higher densities. This is in keeping with the trend of $T_c$ seen experimentally[48] and as shown in the inset of Fig. 1a. We note that the KMK and Eliashberg analyses are expected to overestimate the pairing strength in the overdoped regime so that in a better approximation the collapse of $T_c$ at high $n$ is expected to be more rapid than shown in the inset of Fig. 3, in closer agreement with experiments (Methods)[79,81-84]. Moreover the absolute magnitude of $T_c$, though only estimated to logarithmic accuracy by our calculations (Methods), suggests that the pairing mechanism we have considered here, as defined by Eqs. 1-3, plays an important and likely dominant role in the formation of electron pairs at least near to optimal doping.

The relevance of this pairing mechanism (Eqs. 1-3) is strikingly supported by the correspondence between the predicted and observed rapid collapse of $T_c$ with increasing pressure and hence increasing $1/\varepsilon_0$ (Figs. 1 and 3). For densities in the range $10^{19}$ to $10^{20}$ cm$^{-3}$ where the measurements were performed, $T_c$ tends to peak at the ferroelectric QCP (see Methods).

We have shown that the interaction described in terms of the dielectric function including the effects of fluctuations of the densities of both the ionic and conduction electron systems can lead to a qualitative understanding of the doping and pressure dependence of the superconducting transition temperature of SrTiO$_3$ and related materials in manner similar to that previously proposed by Takada, but including (i) a more tractable model for the interaction, (ii) more transparent discussions of the relevant hybrid longitudinal modes and limitations of the model Eqs. 1-4, and (iii) comparisons with new experiments the results of which we present here. We do not rule out, however, other contributions to the total pairing interaction that may be important particularly well away from optimal doping.

Near to optimal doping the model leads us to expect the occurrence of a superconducting dome in the temperature-pressure phase diagram with a maximum in the vicinity of the ferroelectric quantum critical point and is likely to guide a search for superconductors in other systems. This is at least qualitatively similar to the behaviour observed in the phenomenon of superconductivity on the border, for example, of magnetic long-range order at low temperatures (Fig. 4). The present model, however, predicts that the transition temperature may have a minimum rather than a maximum near to the ferroelectric quantum critical point at doping levels well below optimal (Methods). Thus, we anticipate a dramatic change in form of the pressure dependence of the superconducting dome as a function of carrier density.

The analysis presented here highlights the crucial importance of ionicity in yielding an extraordinarily strong pairing interaction and may help our understanding of other superconductors with electron-polar phonon coupling[85]. We have identified that the predominant contribution to the formation of cooper pairs originates from induced interactions arising from the effects of the coupled conduction-electron and polar phonon system and described here as hybrid longitudinal modes. Such a model may also find applications in superconductors where the conduction electrons are coupled to carriers originating from a separate hole pocket rather than to a separate ionic system such as that recently attempted in low carrier concentration semi-metals[86]. There may exist circumstances under which the advantageous increase in the Fermi energy with increasing carrier density is not offset by a loss in interaction strength between the carriers and the partially screened ions, in the way implied by Eqs. 1-3, potentially leading to pair formation at elevated temperatures. This might be possible in cases in which the form of the interaction and screening are modified such as in doped ferroelectric relaxors[87-90]. It may also be possible in cases where the strong ionic interactions arise not solely at small wavevectors, as is the case in nearly ferroelectric materials, but also at high wavevectors, such as in ferrielectrics, anti-ferroelectrics, multivalley systems, and related materials, where the neutralising effect produced by the doped carriers may be less effective than that operating in electron doped $SrTiO_3$ and $KTaO_3$ superconductors.

# Methods

## Experimental

Incipient ferroelectric $SrTiO_3$ may be made metallic by a number of methods including oxygen reduction and niobium substitution. Electron-doped metallic specimens of $SrNb_xTi_{1-x}O_3$ were obtained from commercial sources with niobium atomic percentages in the range 0.01% to 1%. The samples were cut into parallelepipeds with approximate dimensions 4mm x 1mm x 0.5mm. Low-resistance Ohmic contacts were achieved by cleaning the surfaces using argon-ion plasma etching followed by sputtering of gold contacts on top of a titanium seed-layer on the top surfaces in a standard Hall bar geometry. Charge-carrier concentrations $n$ in the range $1\times10^{18}$ cm$^{-3}$ to $4\times10^{20}$ cm$^{-3}$ were determined for each specimen by measurements of the Hall resistance $R_{xy}$ at liquid-helium temperatures in a field up to 9T. Four-terminal resistance $R_{xx}$ was measured in zero-field for each sample at ambient pressure down to 50mK using an adiabatic demagnetisation refrigerator. The residual resistance ratio, defined as the resistance at room temperature divided by the resistance at 2K, was greater than 600 for all samples. The sample with the highest superconducting transition temperature $T_c = 0.4$K (i.e. the sample with approximately optimal doping $n = 2 \times 10^{19}$ cm$^{-3}$) was selected for the high pressure experiment. To check for repeatability, low-temperature resistivity measurements under hydrostatic pressure were carried out on two different adiabatic demagnetisation refrigerators, one in Cambridge and one in Rio de Janeiro, using two different piston-cylinder clamp cells. In these experiments, hydrostatic pressure was applied to the sample at room temperature using fluorinert (1:1, FC84-FC87) as the pressure transmitting fluid. For each fixed pressure, four-terminal resistance of the sample was measured using a lock-in amplifier and constant current source as a function of temperature down to 50mK. The pressure was determined in the low temperature range by measuring the superconducting transition temperature of a high-purity tin manometer using the calibration provided in ref. ([91]). For all observations of superconductivity in our $SrNb_xTi_{1-x}O_3$ samples we defined $T_c$ to be the temperature at which the resistivity dropped by 10% from its value on entering the superconducting state. Measurements were collected during cooling and heating runs at a rate of approximately 1 K per hour.

## Theoretical

**The hybrid longitudinal frequencies and coupling parameters:**

The hybrid longitudinal mode frequencies, $\omega_\pm(q)$, and coupling functions, $\gamma_\pm(q)$, can be determined in terms of the starting model parameters in $\varepsilon(q, \omega)$ by writing the sum of unity and two resonances on the right-hand side of Eq. 2 as a ratio of two fourth order polynomials, each of which can be factorized in terms of two second order polynomials. Applying the same procedure to $1/\varepsilon(q, \omega)$ expressed as a sum of unity and two resonances (Eqs. 1 and 3) leads to the following closed form expressions for $\omega_\pm(q)$ and $\gamma_\pm(q)$ in terms of the starting model parameters in Eqs. 2a, 2b and 2c

$$\omega_\pm(q) = \tfrac{1}{2}\left(\omega(q)^2 + \Omega(q)^2 + \omega_p^2 + \Omega_p^2\right)$$
$$\pm \left(\tfrac{1}{4}\left(\omega(q)^2 + \Omega(q)^2 + \omega_p^2 + \Omega_p^2\right)^2 - \omega(q)^2 \Omega_p^2 - \Omega(q)^2 \omega_p^2 - \omega(q)^2 \Omega(q)^2\right)^{1/2} \quad (6a)$$

$$\gamma_\pm(q) = (\omega(q)^2 + \Omega(q)^2 - \omega_\pm(q)^2 - \omega(q)^2 \Omega(q)^2 \omega_\pm(q)^{-2})/(\omega_+(q)^2 - \omega_-(q)^2) \quad (6b)$$

**The KMK kernel:**

Each resonance factor in the interaction function $V(q, \omega)$ (Eq. 3)

$$\frac{\omega_\pm(q)^2}{\omega_\pm(q)^2 - \omega^2} \quad (7a)$$

is replaced in the interaction kernel $U(k, k')$ in the BCS gap equation (Eq. 4) by a factor of the form

$$\frac{\omega_\pm(k' - k)^2}{\omega_\pm(k' - k)^2 - \left(\xi(k') - \xi(k)\right)^2} \quad (7b)$$

in the on-shell approximation, or

$$\frac{\omega_\pm(k' - k)}{\omega_\pm(k' - k) + |\xi(k')| + |\xi(k)|} \quad (7c)$$

in the Kirzhnits, Maksimov, and Khomskii (KMK) approximation[70]. Note that in the KMK approximation the kernel falls off with increasing values of the single-particle energies individually and not their difference as in the perhaps unphysical case of the on-shell approximation.

**Parameters used in numerical calculations:**

The calculation presented in Figs. 2 and 3 are based on Eqs. 1-6 and 7c (KMK kernel) in the effective mass approximation in terms of the following fixed parameters determined from independent experiments in the low temperature limit: $\varepsilon_\infty = 5$, $\varepsilon_0 = 2.5 \times 10^4$ [3,92], $m = 4m_e$ [30,61] (relevant near to optimal doping), $\Omega_{LO} = 100$ meV [60] (see caption of Fig. 2), $v_s = 5$ meVÅ [59] and

$$\Omega(0) = \Omega_{LO} \left( \frac{\varepsilon_\infty}{\varepsilon_0} \left( 1 + \frac{n}{\Delta n} \right) \left( 1 + \frac{p}{\Delta p} \right) \right)^{1/2} \quad (8)$$

where $\Delta n = 1.0 \times 10^{19}$ cm$^{-3}$ and $\Delta p = 0.7$ kbar. The solution of the gap equation also involves an upper cut-off wavevector or corresponding frequency taken to be $4\Omega_{LO}$. This cut-off affects the magnitude of $T_c$ or more generally the highest eigenvalue $\Lambda_h$ in Eq. 5, but does not significantly affect the qualitative dependence of $\Lambda_h$ on the carrier density, $n$, or pressure, $p$ (see also the Extended Data section). The increments in the sums defining the gap equation were reduced in size until the results reached essential convergence. Calculations employing the full Lindhard function, and hence dissipation in the single particle continuum, give similar results for the upper eigenvalues where comparisons have been made.

## Acknowledgements

We thank K. Apostolidou, K. Behnia, P. Chandra, A. V. Chubukov, J. R. Cooper, F. Dinola-Neto, C. Durmaz, R. M. Fernandes, M. Fontes, D. E. Khmelnitskii, D. L. Maslov, A. Mello, R. Ospina, J. F. Scott, and C. M. Varma for useful help and discussions. We thank A. Edelman and P. B. Littlewood for discussing with us their description of the electronic properties and superconductivity in $SrTiO_3$ based on the Eliashberg formalism. SER acknowledges support from a CONFAP Newton grant. CE thanks ICAM for financial support. EBS acknowledges support from several grants of FAPERJ and CNPq and in addition a CNPq BP and Emeritus FAPERJ fellowship. SSS acknowledges support from the Increase Competitiveness Program of the Ministry of Education of the Russian Federation under grant number NUST MISiS K2-2017-024. GGL acknowledges support from the Engineering and Physical Sciences Research Council (EPSRC) grant $N^{o.}$ EP/K012894/1 and the CNPq/Science without Borders Program.

## Author Contributions

This work resulted from a collaborative effort of all of the authors each who played an essential role.

## Competing Financial Interests

The authors declare that they have no competing financial interests.

# Figures

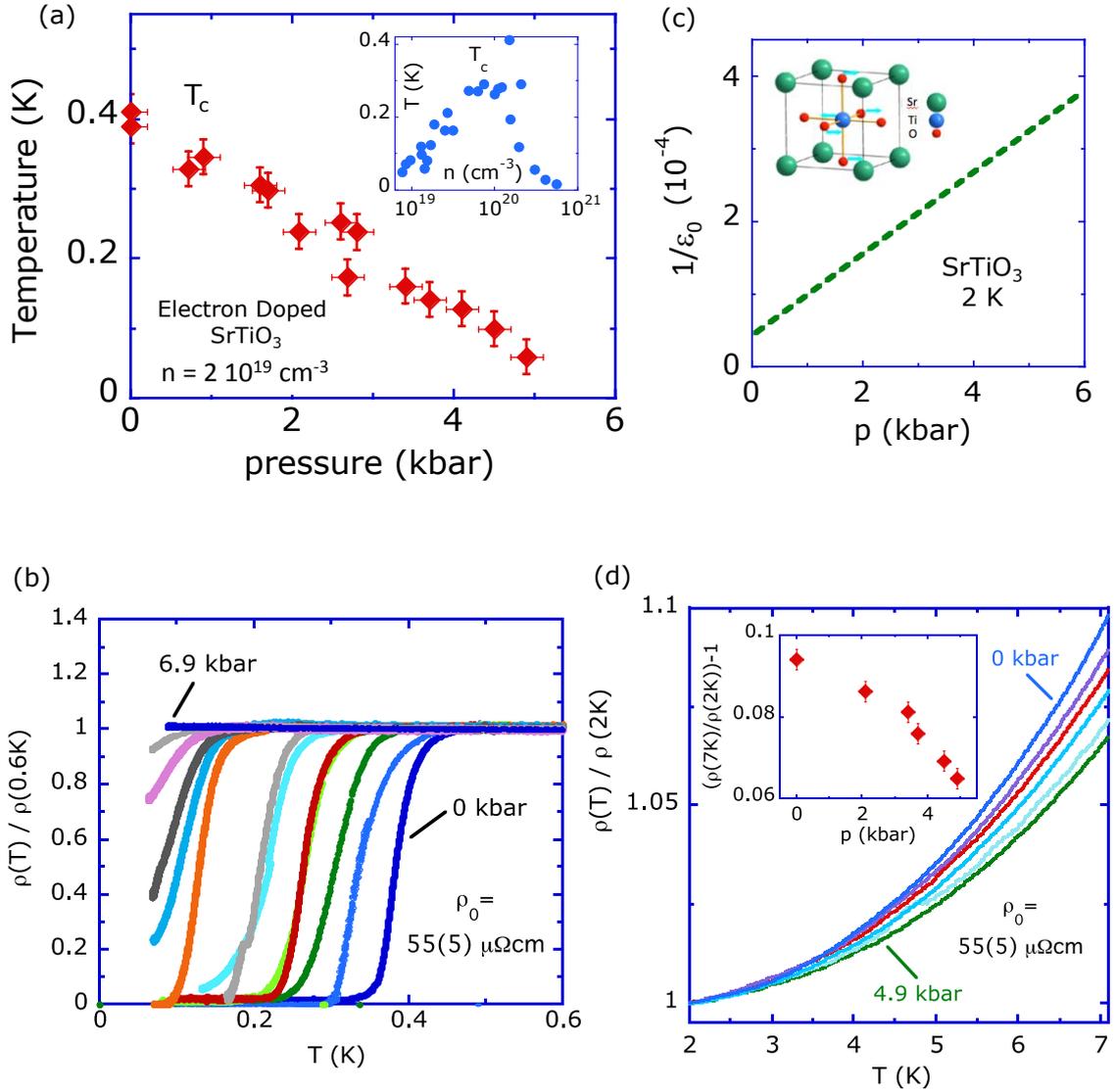

Figure 1. The observed superconductivity temperature-pressure phase diagram for Nb doped $SrTiO_3$. (a) Superconducting transition temperature, $T_c$, for Nb doped $SrTiO_3$ for a carrier density, $n$, near to the dome maximum of the $T_c$ vs. $n$ phase diagram shown in the inset[33] (optimal doping). (b) Temperature dependence of the resistivity below 1K at different applied pressures ($n = 2 \times 10^{19}$ cm$^{-3}$). The resistivity is scaled to the normal state residual value, $\rho_0 = 55(5)$ μΩcm, which is weakly dependent on pressure. The superconducting transition temperatures $T_c$ shown in (a) are determined from the 10% drop of $\rho$ from $\rho_0$. The values of pressure for decreasing values of $T_c$ are $p = 0$, 0 (after decompression), 0.9, 1.6 1.7, 2.6, 2.8, 3.4, 3.7, 4.1, 4.5, 4.9, 6.9 kbar respectively. (c) Pressure dependence of the inverse dielectric constant $1/\varepsilon_0$ (which is proportional to the soft transverse-optical phonon frequency) for the undoped state[5,22,23]. The lowest (soft) transverse-optical phonon frequency is only slightly

changed (see Eq. (8)) up to carrier concentrations of around $10^{19}$ cm$^{-3}$ [60]. (d) Temperature dependence of the resistivity above 1K at different applied pressures ($n = 2 \times 10^{19}$ cm$^{-3}$) scaled to the normal state residual resistivity. The values of pressure may be inferred from the inset, which indicates the variation of the resistivity coefficient $A$ (defined by $\rho = \rho_0 + AT^x$ and in the caption of Fig. S1) vs. pressure. The collapse of $T_c$ with increasing pressure and hence $1/\varepsilon_0$ is seen to be extraordinarily rapid, suggesting that the pairing strength grows dramatically on approaching a ferroelectric quantum critical point (QCP), at least near to the superconducting optimal doping (i. e., for $\omega_p \gg \Omega_p$ as discussed in the text).

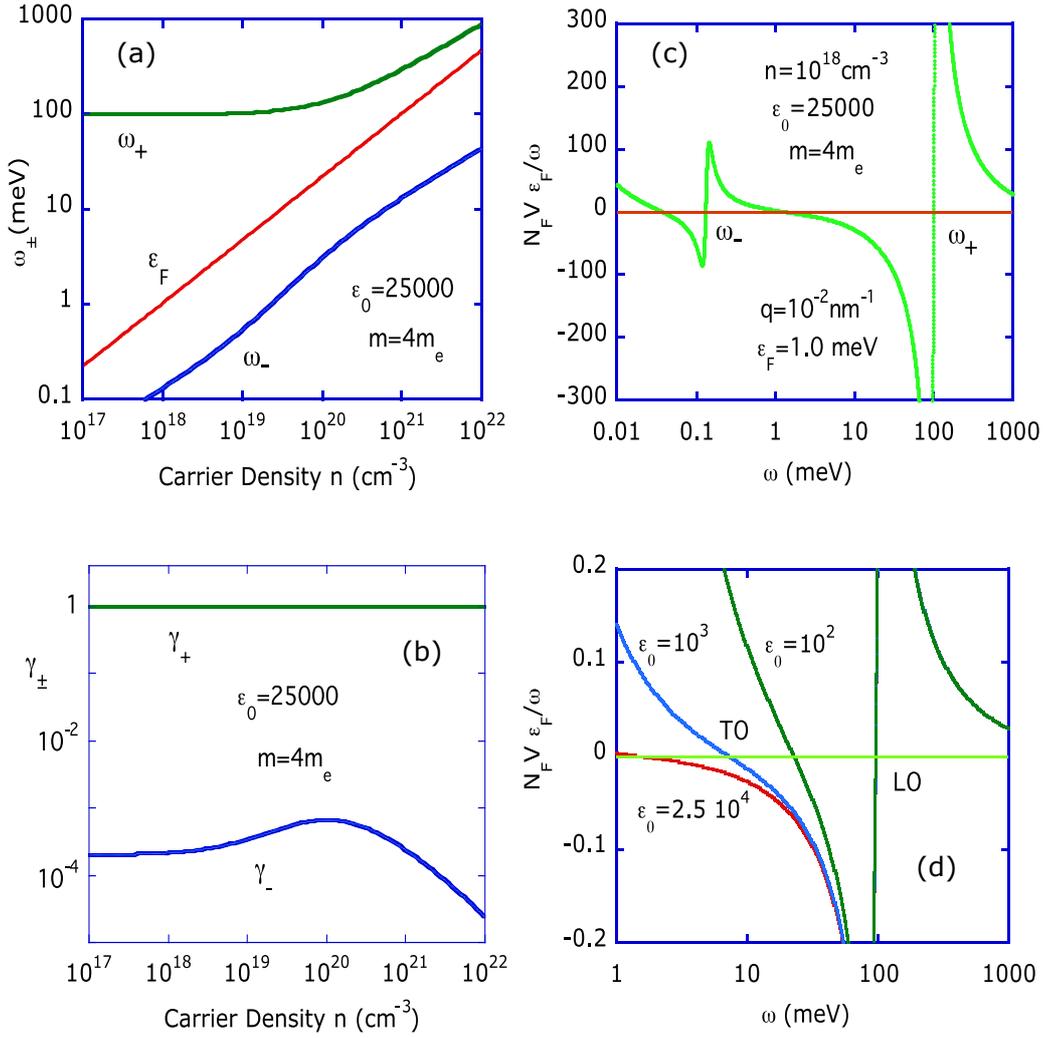

Figure 2. The calculated hybrid longitudinal optical frequencies and effective interaction between doped carriers for Nb doped $SrTiO_3$. (a and b) The hybrid longitudinal frequencies $\omega_\pm(q)$ and coupling functions $\gamma_\pm(q)$ vs. carrier density, $n$, defined by Eqs. 1-3 and materials parameters relevant to $SrTiO_3$ (Methods). Also shown in (a) is the Fermi energy, $\varepsilon_F$. (We adopt units such that $\hbar = k_B = 1$ throughout). (c) The characteristic form of the normalized effective interaction $N_F V(q,\omega)\varepsilon_F/\omega$ vs. $\omega$ for a representative carrier density, $n$, and a low value of $q \ll k_F$, where $N_F$ is the density of states of the doped carriers at the Fermi level and $V(q,\omega)$ is defined by Eqs. 1-3. The TO frequency and effective mass for the lower $t_{2g}$ band, $m$, are set to representative values (Methods). The $n$ dependences of these quantities become significant at high $n$ and are included in the calculations of $T_c$ presented in Fig. 3. (d) The characteristic form of $N_F V(q,\omega)\varepsilon_F/\omega$ vs. $\omega$ in the limit $n$ & $q \to 0$ for different values of $\varepsilon_0$. The range of the attractive interaction increases with increasing $\varepsilon_0$ and extends down to $\omega \to 0$ at the ferroelectric quantum critical point.

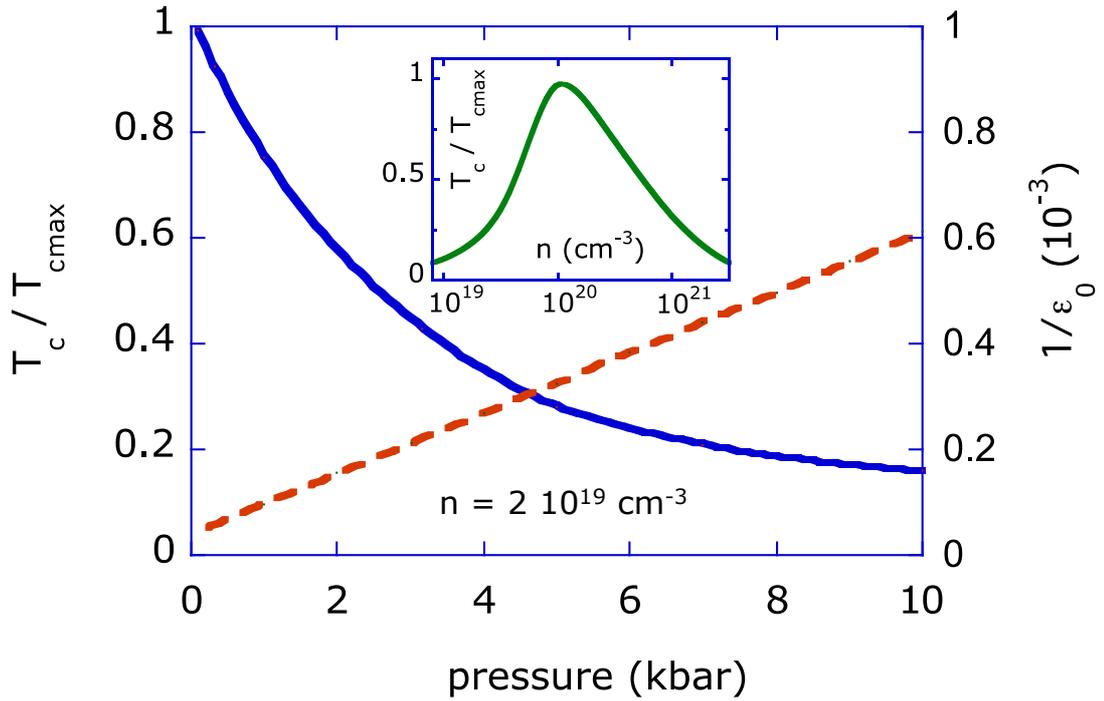

Figure 3. The calculated superconductivity phase diagram for Nb doped $SrTiO_3$. The normalized superconducting transition temperature, $T_c$ (solid line), calculated via the model given by Eqs. 1-4 and the materials parameters relevant to $SrTiO_3$ (Methods), is shown vs. pressure, $p$, near to optimal doping and vs. carrier density, $n$, in the main figure and inset, respectively. $T_{cmax}$ corresponds to the maximum of $T_c$ at $p = 0$ in the main figure and the dome maximum in the inset. Also shown is the pressure dependence of $1/\varepsilon_0$ (dashed line) assumed in the calculations (Fig. 1b)[5,23]. The ratio $T_c/T_{cmax}$ in contrast to $T_c$ itself depends only weakly on the cut-off frequency used to evaluate Eqs. 1-4, and is thus essentially free of adjustable parameters (see also Methods and Extended Data). The close correspondence with the observed phase diagram (Fig. 1a) suggests that the pairing mechanism modelled by Eqs. 1-3 plays a central role in superconductivity in $SrTiO_3$. $T_{cmax}$ is between 0.2K and 0.5K based on a realistic choice of material parameters as outlined in Methods for $SrTiO_3$ and as originating purely from the model described by Eqs. 1-3 in the main text. As discussed in the text and under Methods, the KMK as well as the Eliashberg descriptions are thought to overestimate $T_c$ near and above the dome maximum[81]. In a better approximation for the kernel we expect a more rapid roll off of $T_c$ in the overdoped regime than shown in the inset, in qualitative agreement with observation (Fig. 1a inset).

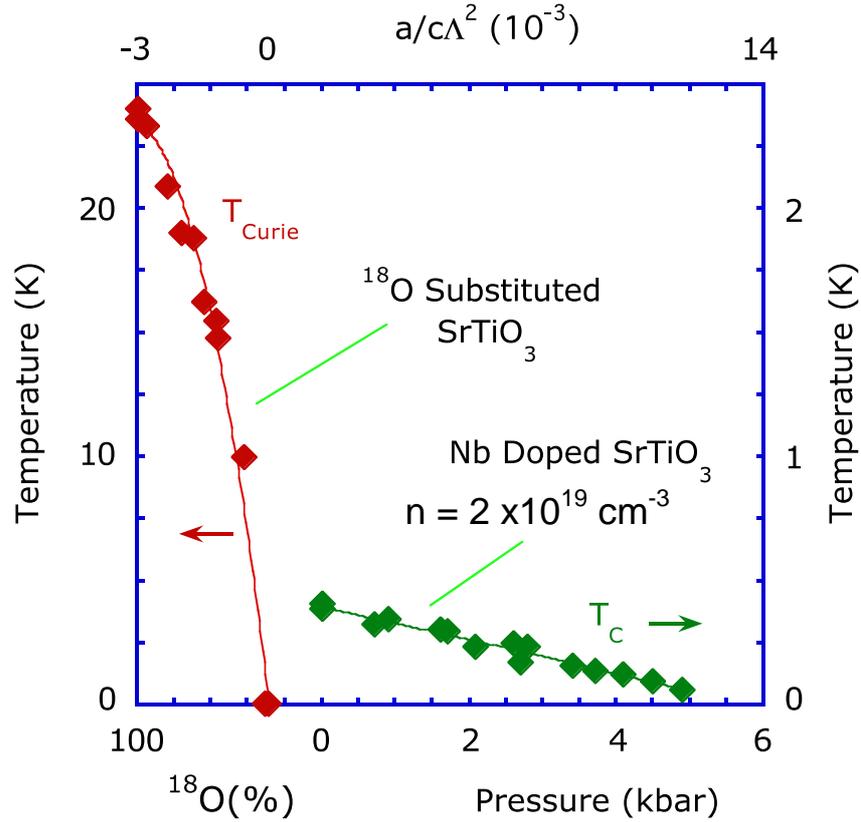

Figure 4. The overall observed phase diagram for SrTiO$_3$ near to the ferroelectric quantum critical point (QCP). The top horizontal axis, or quantum tuning axis, may be defined in terms of the dimensionless parameter given previously[3], which in the present experiments may be calculated from the applied pressure above the QCP and isotopic substitution below (bottom horizontal axis). In general the dimensionless quantum tuning parameter $a/(c\Lambda^2)$ as defined in ref. (3) may be varied by pressure, chemical substitution, strain, isotope substitution and charge-carrier density such that quantum ferroelectrics may be tuned between ferroelectric, paraelectric and superconducting phases. From left to right, the Curie temperature collapses with increasing quantum-tuning parameter and gives way to superconductivity in the presence of carrier doping. Note that at the carrier density, $n = 2 \times 10^{19}$ cm$^{-3}$, given in the figure, the soft transverse-optical frequency and hence proximity to the QCP is essentially the same as that in the undoped state[60] (Methods). This overall phase diagram is consistent with the model presented here (Eqs. 1-4) for the superconducting transition, and with the model presented in ref. (3) for the Curie temperature, both calculated in terms of independently measured (and temperature-independent) model parameters for SrTiO$_3$. These findings, taken together with the results of recent studies of Ca[63] and isotopically doped[64] samples of SrTiO$_3$, suggest that at least for $n$ near optimal doping the superconducting dome along the quantum tuning parameter axis is centred near to the ferroelectric QCP.

**Extended Data**

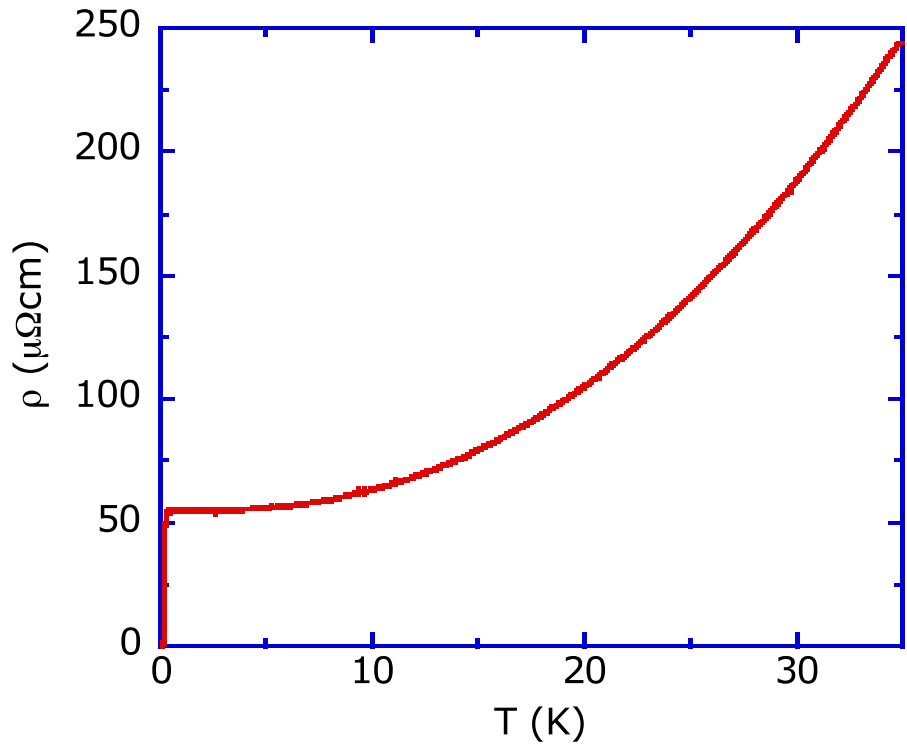

Figure S1. Temperature dependence of the resistivity of Nb doped $SrTiO_3$ with carrier density $n = 2 \times 10^{19}$ cm$^{-3}$. A fit to a function of the form $\rho = \rho_0 + AT^x$ yields an exponent $x$ close to 2. In agreement with previous observations[30,66,67] the exponent is well below that normally expected in terms of the conventional theory of electron phonon scattering.

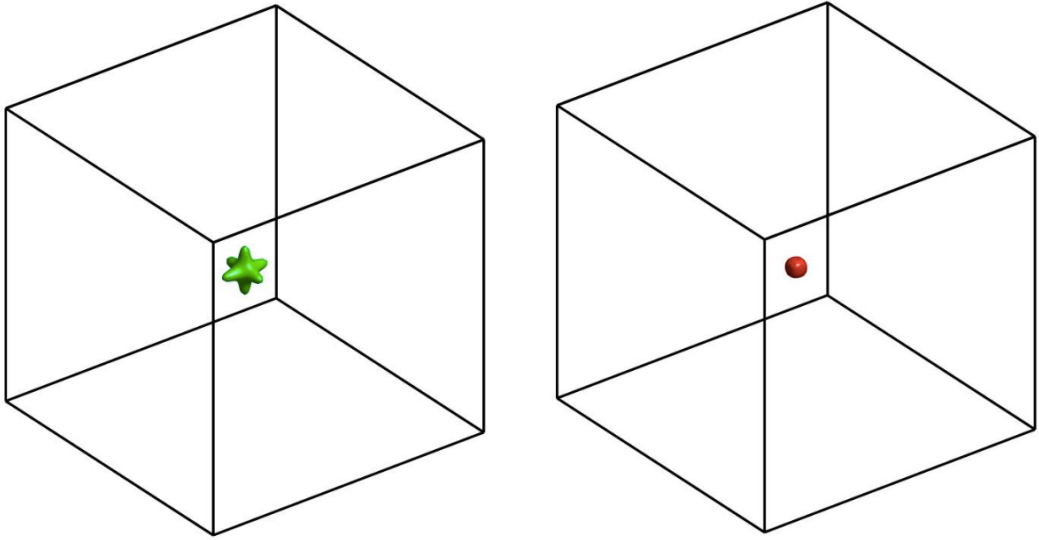

Figure S2. The Fermi surface calculated via Density Functional Theory (DFT) for carrier doped $SrTiO_3$, for *n* of approximately $10^{19}$ cm$^{-3}$. Two of the three t2g bands are populated; the lower band with the fluted Fermi surface dominates the density of states at the Fermi level (see Fig. 2 for the Fermi energy vs. carrier density). The boundaries of the first Brillouin zone of $SrTiO_3$ in its undistorted cubic lattice structure are indicated. The density functional technique employed in the calculations was carried out using the all-electron Wien2k code, with an augmented plane wave (APW) basis using the LDA. Muffin tin radii were set to 2.36, 1.72 and 1.54 a.u. for Sr, Ti and O, respectively, with RKMAX = 7. The ambient pressure experimental lattice parameter was used with 3375 k-points in the full Brillouin zone. Carrier doping was achieved via a rigid shift of the Fermi level.

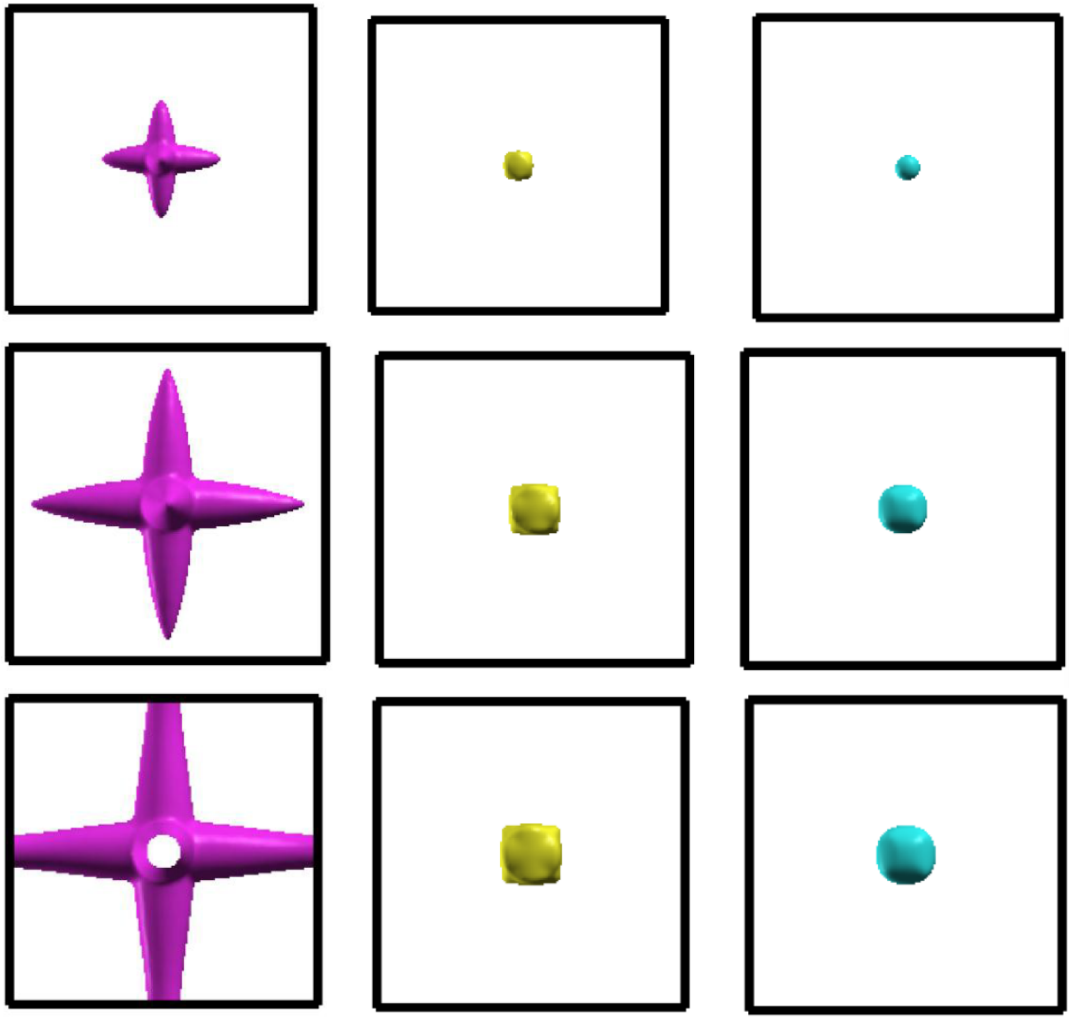

Figure S3. The Fermi surface calculated via Density Functional Theory (DFT) as explained in the caption to Fig. S2 for carrier doped STiO$_3$, for $n$ in the optimally doped to the strongly overdoped regimes from top to bottom ($n\sim 10^{20}$, $10^{21}$, $10^{22}$ cm$^{-3}$, respectively). All three t2g bands are populated. The lower band yielding the Fermi surfaces shown on the left strongly dominates the density of states at the Fermi level (see also Fig. 2).

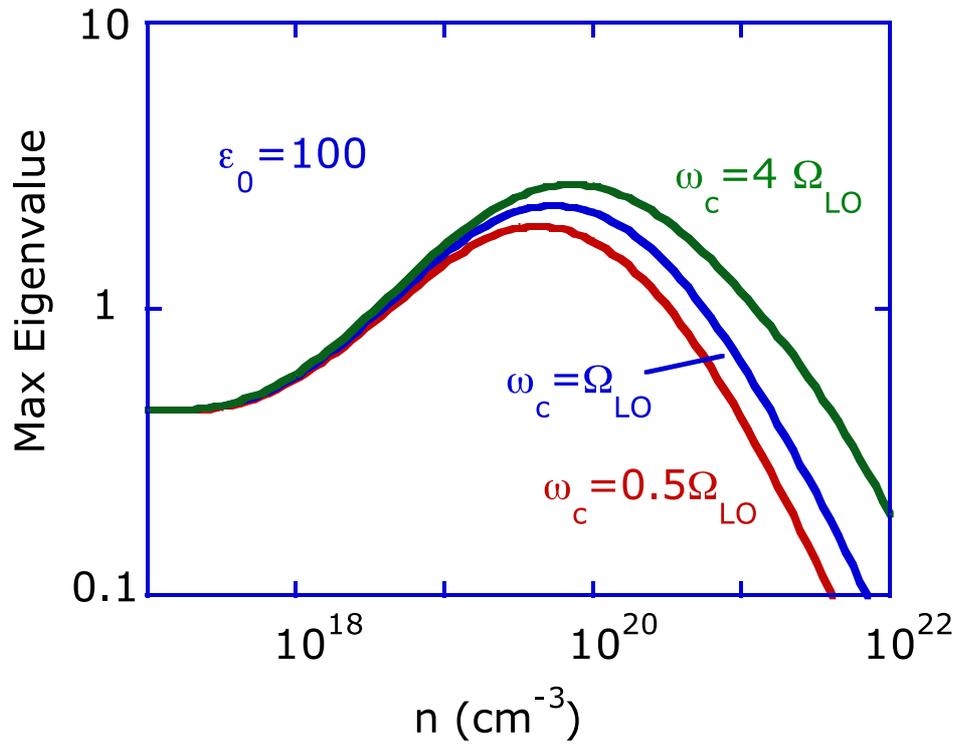

Figure S4. The maximum eigenvalue of the BCS gap equation calculated via the model described in the main text and under Methods for different values of the cut-off frequency $\omega_c$. The parameters entering the calculations are as given under Methods and the calculations are carried out for a reference temperature of $10^{-3}$ K. We note that the forms of the maximum eigenvalues are relatively insensitive to the choice of the cut-off frequency over a sizeable range.

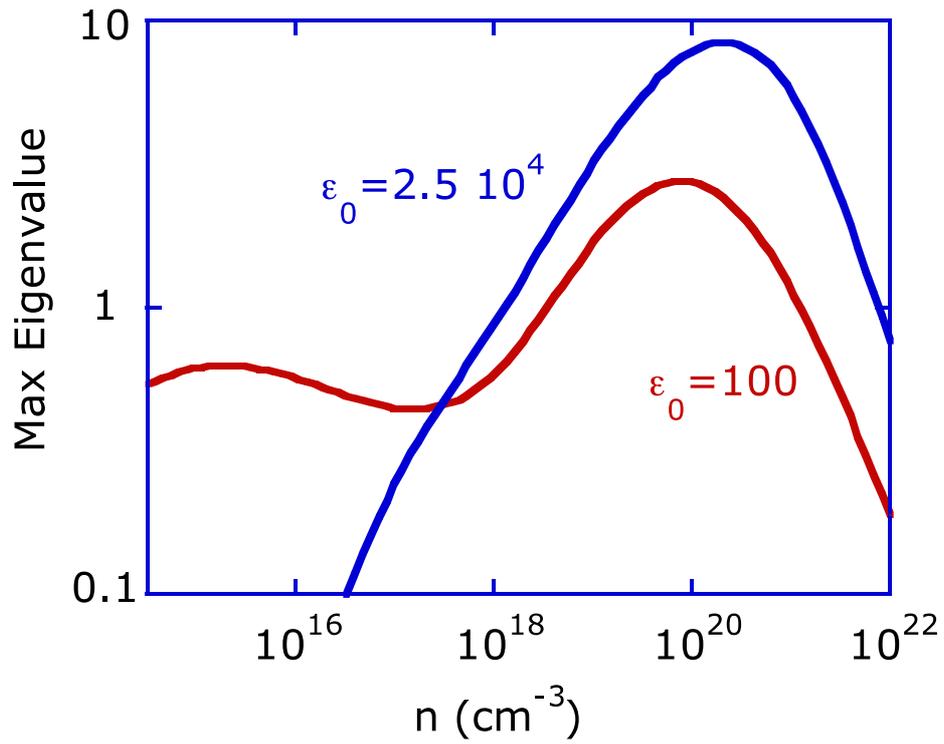

Figure S5. The maximum eigenvalue of the BCS gap equation, calculated for a reference temperature of $10^{-3}$ K, via the model described in the main text and under Methods, except for the different values of $\varepsilon_0$ as labelled in the figure. The maximum eigenvalue increases with increasing $\varepsilon_0$ at high densities, in agreement with observation (Fig. 1). Below of the order of $10^{18}$ cm$^{-3}$, however, the behaviour is inverted, qualitatively as predicted recently by other techniques[50,52].